
%
\input amstex
\documentstyle{amsppt}
\magnification=1200
\NoRunningHeads
\define\Texp{\operatorname{Texp}}
\topmatter
\title THE ADVANTAGE OF A MULTI--USER MODE
\endtitle
\author Denis JURIEV
\endauthor
\dedicatory In memory of my visits to Kiev,\\
"The Mother of Russian Cities".
\enddedicatory
\address Mathematical Division, Research Institute for System Studies
[Information Technologies], Russian Academy of Sciences, Moscow,
Russia\newline
\endaddress
\email juriev\@systud.msk.su
\endemail
\endtopmatter
\document

\

\

\

The purpose of this very short note, which maybe considered as a comment on
[1], is to prove the following proposition.

\proclaim{Proposition} There exist certain models of Mobilevision, in which
interpretational figures are observable only in a multi--user mode in contrast
to single--user ones.
\endproclaim

\demo{Proof}
Let us consider the system of two observers; the dynamical laws of perspective
in MV will be defined by the Euler formulas only, an angular field will
contain only a term proportional to a non--local $q_R$--conformal current
$J(u)$:

$$\dot\Phi=J(u)\dot u\Phi+J(v)\dot v\Phi.$$

In a single--user mode these equations will have the form

$$\dot\Phi=J(u)\dot u\Phi.$$

Such equations maybe integrated:

$$\Phi_u=\Texp\int_0^uJ(u)\,du\cdot\Phi_0,$$
so the observed picture depends only on a position of the sight fixing point.

On the contrary, in the multi--user mode the dynamical equations can't be
split and integrated, so the observed picture is essentially dynamical and is
defined also by the velocities of eye movements.

Let us emphasize that hence the multi--user and single--user modes are
essentially different. Indeed, in the second case a picture maybe defined
before the beginning of any observation process and, therefore, the
interpretation process of an observation has equivalent compilation one. On
the contrary, the process of observation in a multi--user mode is essentially
interpretational. \pagebreak

Therefore, the interpretational figures are not observed in a single--user
modes and are observable in multi--user one \qed
\enddemo

This result maybe interesting as for applications as for the understanding of
theoretical foundations of quantum interactive processes (cf. [2]).

\Refs
\roster
\item"[1]" Juriev D., Visualizing 2D quantum field theory: geometry and
informatics of Mobilevision; hep-th/9401067.
\item"[2]" Rovelli C., On quantum mechanics; hep-th/9403015.
\endroster
\endRefs
\enddocument